\documentclass[
aps,
prl,
reprint,
10pt,
preprintnumbers,
nofootinbib,
superscriptaddress,
floatfix
]{revtex4-2}

\usepackage{amsmath}
\usepackage{amssymb}
\usepackage{graphicx}
\usepackage{mathrsfs}
\usepackage{slashed}
\usepackage{array}
\usepackage[dvipsnames]{xcolor}
\usepackage{indentfirst}
\usepackage{multirow}
\usepackage[caption=false]{subfig}
\usepackage{placeins}

\usepackage[colorlinks=true, allcolors=blue]{hyperref}
\usepackage{cleveref}
\usepackage{subfiles}

\newcommand{\rme}{\mathrm{e}}
\newcommand{\rmi}{\mathrm{i}}
\newcommand{\rmd}{\mathrm{d}}

\begin{document}

%%%%%%%%%%%%%%%%%%%%%%%%%%%%%%%%%%%%%%%%%%%%%%%%%%%%%%%%%%%%%%%%%%%%
%%%%%%%%%%%%%%%%%%%%%%%%%%%%%%%%%%%%%%%%%%%%%%%%%%%%%%%%%%%%%%%%%%%%

\title{Complete Access to Leading-Twist $\Lambda$-Baryon Light-Cone Distribution Amplitudes from Lattice QCD}

\collaboration{\bf{Lattice Parton Collaboration ($\rm {\bf LPC}$)}}

\author{\includegraphics[scale=0.10]{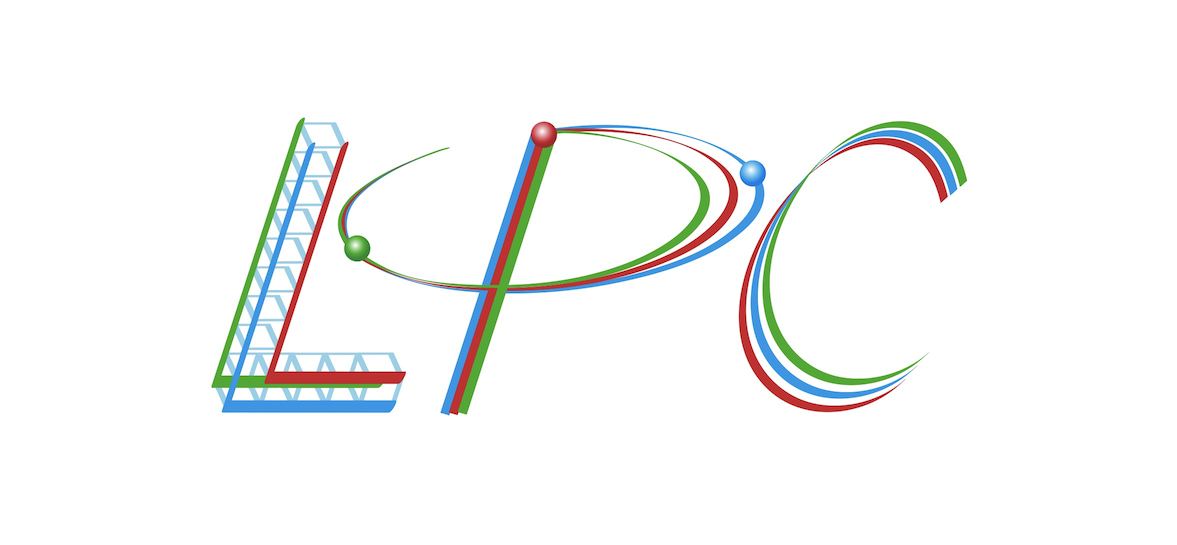}\\Mu-Hua~Zhang}
\affiliation{State Key Laboratory of Dark Matter Physics, Key Laboratory for Particle Astrophysics and Cosmology (MOE), Shanghai Key Laboratory for Particle Physics and Cosmology, Tsung-Dao Lee Institute and School of Physics and Astronomy, Shanghai Jiao Tong University, Shanghai 200240, China}

\author{Haoyang~Bai}
\affiliation{Institute of High Energy Physics, CAS, Beijing 100049, China}
\affiliation{School of Physics, University of Chinese Academy of Sciences, Beijing 100049, China}

\author{Min-Huan~Chu}
\affiliation{Faculty of Physics and Astronomy, Adam Mickiewicz University, ul.\ Uniwersytetu Pozna\'nskiego 2, 61-614 Pozna\'{n}, Poland}

\author{Jun~Hua}
\thanks{Corresponding author.}
\email{junhua@scnu.edu.cn}
\affiliation{State Key Laboratory of Nuclear Physics and Technology, Institute of Quantum Matter, South China Normal University, Guangzhou 510006, China}
\affiliation{Guangdong Basic Research Center of Excellence for Structure and Fundamental Interactions of Matter, Guangdong Provincial Key Laboratory of Nuclear Science, Guangzhou 510006, China}

\author{Xiangdong~Ji}
\affiliation{Tsung-Dao Lee Institute and School of Physics and Astronomy,
Shanghai Jiao Tong University, Shanghai 201210, China}

\author{Xiangyu~Jiang}
\affiliation{CAS Key Laboratory of Theoretical Physics, Institute of Theoretical Physics, Chinese Academy of Sciences, Beijing 100190, China}

\author{Jian~Liang}
\affiliation{State Key Laboratory of Nuclear Physics and Technology, Institute of Quantum Matter, South China Normal University, Guangzhou 510006, China}
\affiliation{Guangdong Basic Research Center of Excellence for Structure and Fundamental Interactions of Matter, Guangdong Provincial Key Laboratory of Nuclear Science, Guangzhou 510006, China}

\author{Cai-Dian~L\"u}
\affiliation{Institute of High Energy Physics, CAS, Beijing 100049, China}
\affiliation{School of Physics, University of Chinese Academy of Sciences, Beijing 100049, China}

\author{Andreas~Sch\"afer}
\affiliation{Institut f\"ur Theoretische Physik, Universit\"at Regensburg, D-93040 Regensburg, Germany}

\author{Wei~Wang}
\thanks{Corresponding author.}
\email{wei.wang@sjtu.edu.cn}
\affiliation{State Key Laboratory of Dark Matter Physics, Key Laboratory for Particle Astrophysics and Cosmology (MOE), Shanghai Key Laboratory for Particle Physics and Cosmology, School of Physics and Astronomy, Shanghai Jiao Tong University, Shanghai 200240, China}
\affiliation{Southern Center for Nuclear-Science Theory (SCNT), Institute of Modern Physics, Chinese Academy of Sciences, Huizhou 516000, Guangdong Province, China}

\author{Yi-Bo~Yang}
\affiliation{CAS Key Laboratory of Theoretical Physics, Institute of Theoretical Physics, Chinese Academy of Sciences, Beijing 100190, China}
\affiliation{School of Fundamental Physics and Mathematical Sciences, Hangzhou Institute for Advanced Study, UCAS, Hangzhou 310024, China}
\affiliation{International Centre for Theoretical Physics Asia-Pacific, Beijing/Hangzhou, China}
\affiliation{School of Physical Sciences, University of Chinese Academy of Sciences,
Beijing 100049, China}

\author{Jian-Hui~Zhang}
\affiliation{School of Science and Engineering, The Chinese University of Hong Kong, Shenzhen 518172, China}

\author{Jia-Lu~Zhang}
\affiliation{State Key Laboratory of Dark Matter Physics, Key Laboratory for Particle Astrophysics and Cosmology (MOE), Shanghai Key Laboratory for Particle Physics and Cosmology, Tsung-Dao Lee Institute and School of Physics and Astronomy, Shanghai Jiao Tong University, Shanghai 200240, China}

\author{Qi-An~Zhang}
\affiliation{School of Physics, Beihang University, Beijing 102206, China}

\begin{abstract}
We report the first complete lattice-QCD determination of the leading-twist light-cone distribution amplitudes (LCDAs) of the $\Lambda$ baryon, obtained as full two-dimensional functions of the valence-quark momentum fractions. The calculation employs large-momentum effective theory to relate the light-cone amplitudes to equal-time nonlocal three-quark matrix elements of boosted $\Lambda$ baryons. Controlled extrapolations to the continuum, physical-pion-mass, and infinite-momentum limits, together with state-of-the-art lattice techniques including hybrid renormalization, large-$\lambda$ extrapolation, and perturbative matching, yield the three leading-twist LCDAs $V$, $A$, and $T$ at physical point. Using the lattice-determined LCDAs in place of the asymptotic form, we find a $\mathcal{O}(10\%)$ shift in the $\Lambda$ electromagnetic form factor at perturbative scales, demonstrating that the full $x$-dependent LCDAs, rather than only their asymptotic shapes or lowest moments, are required for precision baryonic phenomenology. This work, together with the companion article~\cite{LPC:2026mvw} detailing the baryon-LaMET framework, provides the first complete $x$-dependent baryon LCDAs from first principles and establishes a benchmark for lattice access to multi-dimensional baryon structure.
\end{abstract}

\maketitle

%%%%%%%%%%%%%%%%%%%%%%%%%%%%%%%%%%%%%%%%%%%%%%%%%%%%%%%%%%%%%%%%%%%%
%%%%%%%%%%%%%%%%%%%%%%%%%%%%%%%%%%%%%%%%%%%%%%%%%%%%%%%%%%%%%%%%%%%%
\paragraph{Introduction.} The $\Lambda$ baryon occupies a distinctive position in hadron physics. As the lightest strange baryon, its parity-violating weak decay $\Lambda\to p\pi^-$ makes the angular distribution of its decay products a self-analyzing probe of hyperon polarization and discrete symmetries~\cite{Lee:1956qn,BESIII:2018cnd}. More broadly, the first observation of CP violation in baryon decays by the LHCb Collaboration~\cite{LHCb:2025ray} has underscored the emergence of baryons as a new precision frontier in flavor and CP physics. For the $\Lambda$ itself, increasingly precise measurements of its timelike electromagnetic form factors probe its electromagnetic structure~\cite{BESIII:2025kyg,BESIII:2025yzk}, while searches for its electric dipole moment (EDM) test possible CP-violating interactions. Using quantum-entangled $\Lambda\bar{\Lambda}$ pairs, a recent BESIII analysis obtained a result consistent with zero and improved the sensitivity to the $\Lambda$ EDM by nearly three orders of magnitude, with the same method applicable to other members of the hyperon family~\cite{BESIII:2025vxm}. 
Together, these developments make the $\Lambda$ a timely target for a first-principles description of its short-distance three-quark structure.

A basic set of nonperturbative quantities characterizing this short-distance structure is provided by the leading-twist baryon light-cone distribution amplitudes (LCDAs). At sufficiently large momentum transfer, they describe the longitudinal momentum structure of the valence three-quark Fock component and enter QCD factorization formulas for hard exclusive baryonic processes~\cite{Lepage:1980fj,Chernyak:1987nu}.
Unlike meson LCDAs, which depend on only one independent momentum fraction, baryon LCDAs depend on two and are therefore genuinely two-dimensional distributions. 
They resolve the correlated sharing of longitudinal momentum among three valence quarks, together with the associated spin-flavor and permutation structures. 
This makes baryon LCDAs substantially richer than their mesonic counterparts~\cite{Chernyak:1987nu,Braun:1999te}.

For the $\Lambda$, resolving the complete momentum-fraction dependence is especially important.
With an isosinglet light-quark pair and a strange valence quark, the $\Lambda$ provides a natural system for investigating flavor-dependent momentum sharing within a baryon.
The leading-twist $\Lambda$ LCDAs enter short-distance descriptions of its electromagnetic form factors, while baryon LCDAs more broadly provide nonperturbative inputs to weak-decay amplitudes and CP asymmetries~\cite{Han:2024kgz,Han:2025tvc}.
More recently, a perturbative-QCD analysis related the CP-odd electric dipole form factor of the $\Lambda$ to LCDA convolutions, connecting hyperon EDM searches to quark electric and chromoelectric dipole interactions, with distinctive sensitivity to the strange-quark chromoelectric dipole moment~\cite{Chen:2025rab}.
The $\Lambda$ also presents a distinctive structural challenge: its leading-twist $V$ and $T$ amplitudes are antisymmetric under interchange of the two light quarks and consequently have vanishing zeroth moments.
Although their nonlocal distributions are nonzero, they vanish in the local limit and cannot be characterized by independent leading-twist decay constants.
A direct nonlocal determination of their full two-dimensional momentum dependence is therefore indispensable.

Existing information on $\Lambda$ LCDAs has mainly come from QCD sum rules and OPE-based lattice calculations of normalization constants and first moments~\cite{Chernyak:1984bm,King:1986wi,Liu:2014uha,Braun:2006hz,Bali:2024oxg}. Such information does not determine the complete two-dimensional distributions required in hard-scattering convolutions. Large-momentum effective theory (LaMET)~\cite{Ji:2013dva,Ji:2020ect} provides a first-principles route to light-cone correlations through equal-time spatial matrix elements of boosted hadrons. Its extension from mesons to baryons, however, introduces qualitatively new challenges: nonlocal three-quark operators depend on two spatial separations and require genuinely two-dimensional renormalization, large-$\lambda$ extrapolation, and perturbative matching. The corresponding factorization and renormalization framework has recently been established~\cite{Deng:2023csv,Han:2023xbl,Han:2024ucv,Zhang:2025npd}, while previous numerical studies have demonstrated the feasibility of the baryon-LaMET construction and its hybrid renormalization~\cite{LatticeParton:2024vck,LatticePartonCollaborationLPC:2025vhd}.

In this Letter, we present the first lattice-QCD determination of the complete set of leading-twist $\Lambda$-baryon LCDAs, $V$, $A$, and $T$, as full two-dimensional functions of the independent valence-quark momentum fractions. 
Combining boosted nonlocal three-quark matrix elements with hybrid renormalization, large-$\lambda$ extrapolation, perturbative matching, and controlled extrapolations to the continuum, physical-pion-mass, and infinite-momentum limits, we directly resolve both the dominant symmetric amplitude and the antisymmetric structures that vanish in the local limit. 
The resulting amplitudes reveal the correlated momentum sharing among the light and strange valence quarks and provide first-principles nonperturbative inputs for hard exclusive processes and future analyses of CP-sensitive hyperon form factors. 
As a representative phenomenological test, replacing the commonly used asymptotic form by the lattice-determined LCDAs changes the perturbative $\Lambda$ Dirac form factor by approximately $12\%$ at next-to-leading order. 
This work therefore provides the first first-principles resolution of the genuinely multi-dimensional light-cone structure of a baryon and demonstrates the quantitative sensitivity of a hard exclusive observable to its full two-dimensional LCDA shape.

%%%%%%%%%%%%%%%%%%%%%%%%%%%%%%%%%%%%%%%%%%%%%%%%%%%%%%%%%%%%%%%%%%%%
%%%%%%%%%%%%%%%%%%%%%%%%%%%%%%%%%%%%%%%%%%%%%%%%%%%%%%%%%%%%%%%%%%%%
\paragraph{$\Lambda$ LCDAs within LaMET.} The baryon LCDAs are defined from gauge-invariant baryon-to-vacuum matrix elements of three light-cone-separated quark fields~\cite{Braun:1999te,Han:2024ucv}. At leading twist, the matrix elements involve three Dirac-$\gamma$ structures $V$, $A$, and $T$, with the corresponding coordinate-space amplitudes denoted by $\Phi_\Lambda^{V,A,T}$. In this work we adopt the convention $\xi_3=0$ and define the momentum-space amplitudes as:
\begin{equation}
\begin{aligned}
    \phi_\Lambda^X(x_1,x_2;\mu)
    =
    &\ (P^+)^2\int\frac{\rmd\xi_1}{2\pi}\frac{\rmd\xi_2}{2\pi}\rme^{\rmi(x_1\xi_1+x_2\xi_2)P^+}\\
    &\ \times\Phi_\Lambda^X(\xi_1P^+,\xi_2P^+;\mu)\ ,
\end{aligned}
\end{equation}
where $P^+=n\cdot P$, $X=V,A,T$, and the physical support is the triangular region $0\le x_1,x_2,x_3\le 1$ and $x_1+x_2+x_3=1$. This definition makes explicit that, in contrast to meson LCDAs, leading-twist baryon LCDAs are genuinely two-dimensional functions of the valence-quark momentum fractions.

\begin{figure}[htbp]
    \centering
    \includegraphics[width=0.9\linewidth]{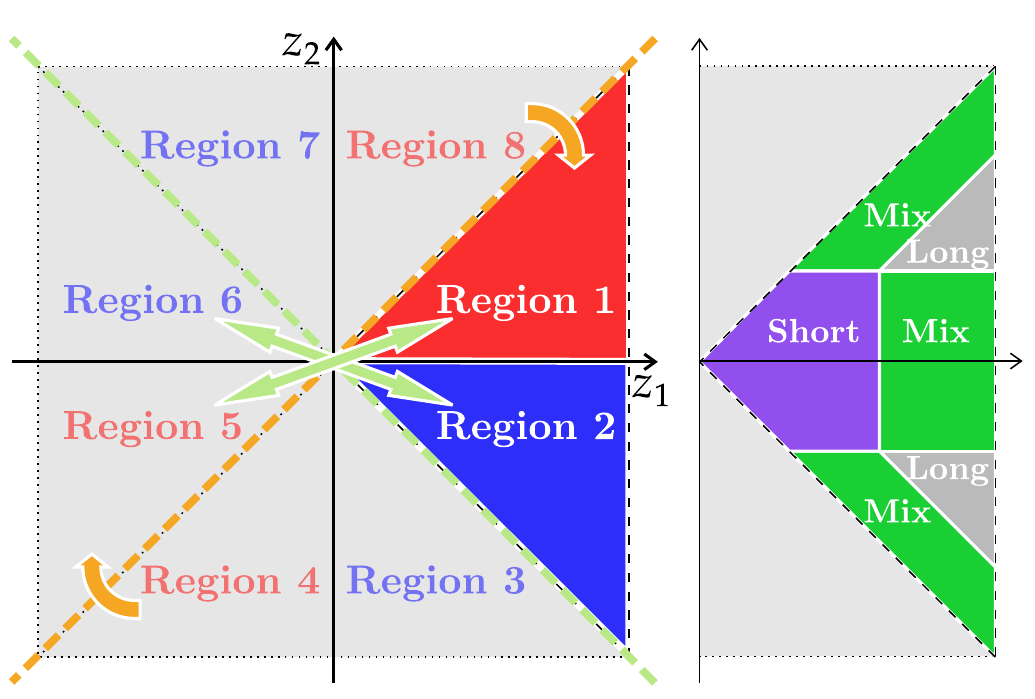}
    \caption{Region partition in the $(z_1,z_2)$ plane. In left panel, the gray regions are related to regions 1 and 2 by the symmetries of $\Lambda$ quasi-DAs. The right panel shows the partition in independent regions according to the short- and long-distance behavior of the three separations $|z_1|$, $|z_2|$, and $|z_1-z_2|$. Further details are given in the companion article~\cite{LPC:2026mvw}.}
    \label{fig:regions}
\end{figure}

For the $\Lambda$ baryon, the LCDAs are constrained by the exchange symmetries and the reality condition:
\begin{equation}
\begin{aligned}
    \phi_\Lambda^A(x_1,x_2;\mu)&=+\phi_\Lambda^A(x_2,x_1;\mu)\ , \\
    \phi_\Lambda^{V,T}(x_1,x_2;\mu)&=-\phi_\Lambda^{V,T}(x_2,x_1;\mu)\ , \\
    \phi_\Lambda^{V,A,T}(x_1,x_2;\mu)&={\rm Re}\ \phi_\Lambda^{V,A,T}(x_1,x_2;\mu)\ .
\end{aligned}
\end{equation}
These properties imply that only two regions in the coordinate $(z_1,z_2)$ plane contain independent data, as illustrated in Fig.~\ref{fig:regions}, while the remaining regions can be reconstructed by symmetries. They also lead to the normalization conditions:
\begin{equation}
    \int[\rmd x] \phi_\Lambda^A(x_1,x_2;\mu)=1\ ,\ 
    \int[\rmd x] \phi_\Lambda^{V,T}(x_1,x_2;\mu)=0\ ,
\end{equation}
where $\int[\rmd x]\equiv \int_0^1 \rmd x_1\int_0^{1-x_1}\rmd x_2$. This indicates that only the $A$ amplitude has a non-vanishing local value, whereas the antisymmetric $V$ and $T$ amplitudes vanish in the local limit.
This feature is central to the lattice construction: the $V$ and $T$ quasi-DAs cannot be normalized by their own local matrix elements; instead, their nonlocal correlators are reduced using the nonzero local $A$ correlator.
These symmetries also constrain the choice of reference matrix elements in the following hybrid renormalization procedure.

%%%%%%%%%%%%%%%%%%%%%%%%%%%%%%%%%%%%%%%%%%%%%%%%%%%%%%%%%%%%%%%%%%%%
%%%%%%%%%%%%%%%%%%%%%%%%%%%%%%%%%%%%%%%%%%%%%%%%%%%%%%%%%%%%%%%%%%%%

Since the LCDAs are defined through light-cone correlators, they are not directly accessible in Euclidean lattice QCD. In LaMET, one instead computes equal-time quasi-DAs for a baryon boosted with a large but finite momentum $P^z$, and relates them to the physical LCDAs through the LaMET matching formula:
\begin{equation}\label{eq:LaMET}
\begin{aligned}
    \phi^X_B(x_1,x_2;\mu)
    =
    \int&\ \rmd y_1\rmd y_2\ \mathcal C^X(x_1,x_2;y_1,y_2;P^z,\mu)\\
    &\ \qquad\times\widetilde \phi^X_B(y_1,y_2;P^z,\mu)\\
    + \mathcal O\Bigg( \frac{\Lambda_{\rm QCD}^2}{(x_1P^z)^2}, &\frac{\Lambda_{\rm QCD}^2}{(x_2P^z)^2}, \frac{\Lambda_{\rm QCD}^2}{[(1-x_1-x_2)P^z]^2} \Bigg)\ .
\end{aligned}
\end{equation}
The coordinate-space quasi-DAs are defined from equal-time three-quark matrix elements with spatial separations along the boost direction $n_z^\mu=(0,0,0,1)$.
For the $\Lambda$ baryon, they are obtained schematically from
\begin{equation}
\begin{aligned}
    &\ \widetilde{\Phi}_{\Lambda}^X(z_1,z_2;P^z,\mu) \ \propto \ \widetilde{M}_{\Lambda}^X(z_1,z_2;P^z,\mu)\\
    =&\ \langle 0|u^{\rm T}(z_1n_z)\Gamma^X_1 d(z_2n_z) \Gamma^X_2 s(0) |B(P^z)\rangle^R\ ,
\end{aligned}
\end{equation}
where the choices of $\Gamma^X_1$ and $\Gamma^X_2$ project the $V$, $A$, and $T$ amplitudes. 
The lattice calculation therefore first gives the coordinate-space quasi-DAs $\widetilde{\Phi}_\Lambda^X(z_1,z_2;P^z,\mu)$, while the momentum-space quasi-DAs $\widetilde{\phi}_\Lambda^X(y_1,y_2;P^z,\mu)$ are obtained by a two-dimensional Fourier transform. 

The bare coordinate-space quasi-DAs contain ultraviolet divergences, including the Wilson-line linear divergence.
We renormalize them using a hybrid prescription on the two-dimensional $(z_1,z_2)$ plane.
At short distances, a ratio scheme is used to cancel the ultraviolet dependence through division by zero-momentum matrix elements.
At long distances, the self-renormalization is used instead, avoiding uncontrollable infrared contamination from long-distance structures.
The two prescriptions are combined according to the three relevant separations $|z_1|$, $|z_2|$, and $|z_1-z_2|$, as illustrated in Fig.~\ref{fig:regions}.
For the antisymmetric $V$ and $T$ amplitudes, the reference matrix elements are chosen consistently with the symmetries discussed above.
The one-loop LaMET matching is then implemented in the same hybrid scheme, with the corresponding hybrid counterterms included in the matching kernel.

The lattice data for the quasi-DAs are available only in a finite region of the coordinate $(z_1,z_2)$ plane, whereas the momentum-space quasi-DAs entering the LaMET matching require a two-dimensional Fourier transform.
A direct transform of the finite-range coordinate-space data would therefore introduce sizable finite-range artifacts, especially near the endpoint regions.
To control this effect, we extend the coordinate-space matrix elements to larger quasi light-cone distances $\lambda=zP^z$ using the asymptotic analysis of Euclidean correlators newly developed in Ref.~\cite{Ji:2026vir}.
For a single spatial separation, the large-distance behavior takes the form:
\begin{align} \label{eq:asym_formula}
    &\ \widetilde{\Phi}\left(z; P^z\right)=\sum_{\Lambda^{J^P}} \rme^{-\Lambda^{J^P}|z|} \\
    &\times\left[ \left( \mathcal H_1 \rme^{-\rmi z P^z} + \mathcal H_2 \right) + \left( \mathcal H_1^\prime \rme^{-\rmi z P^z} + \mathcal H_2^\prime\right) \frac{1}{|z|} + \cdots \right]\ ,\notag
\end{align}
where $\Lambda^{J^P}$ denotes the binding energy of intermediate state associated with the HQET description of the Wilson line, and $\mathcal H_n^{(\prime)}$ denotes the accuracy of asymptotic large-$|z|$ expansion.
We generalize this construction to baryon quasi-DAs, for which the coordinate-space matrix elements depend on three large-distance variables $|z_1|$, $|z_2|$, and $|z_1-z_2|$.
The $(z_1,z_2)$ plane is therefore partitioned again, according to which separation becomes large, and the resulting baryon-specific asymptotic forms are used as the extrapolation ans\"atze in each region.
This procedure provides a smooth coordinate-space completion, with the difference between the leading- and next-to-leading-asymptotic forms in the $|z|$-expansion included as a systematic uncertainty when both fits are stable.

The calculation proceeds from equal-time coordinate-space quasi-DAs of boosted $\Lambda$ baryon to physical LCDAs through two-dimensional hybrid renormalization, coordinate-space large-$\lambda$ extrapolation, Fourier transform, and LaMET matching.
The symmetries of the $\Lambda$ baryon constrain the treatment of the antisymmetric $V$ and $T$ amplitudes both in the reduction of lattice correlators and in the choice of renormalization references.
The detailed operator definitions, reduction formulas, hybrid renormalization, and baryon-specific asymptotic forms are given in the companion article~\cite{LPC:2026mvw}.

%%%%%%%%%%%%%%%%%%%%%%%%%%%%%%%%%%%%%%%%%%%%%%%%%%%%%%%%%%%%%%%%%%%%
%%%%%%%%%%%%%%%%%%%%%%%%%%%%%%%%%%%%%%%%%%%%%%%%%%%%%%%%%%%%%%%%%%%%
\paragraph{Lattice Simulation.---}
The numerical calculation is performed on seven $N_f=2+1$ lattice ensembles generated by the CLQCD Collaboration, using stout-smeared clover fermions coupled with the Symanzik-improved gauge action~\cite{CLQCD:2023sdb,CLQCD:2024yyn}. The ensembles cover pion masses from $m_\pi\approx 317$ to $136~{\rm MeV}$ and lattice spacings from $a\approx 0.105$ to $0.052~{\rm fm}$, while the boosted $\Lambda$ baryon is computed with several momenta up to about $P^z\approx 3~{\rm GeV}$, as summarized in Table~\ref{tab:Ensembles}. 
This setup of ensembles provides the lever arm needed to control the continuum, physical-pion-mass, and infinite-momentum extrapolation to the physical point. 
To improve the signal for boosted baryon matrix elements, we employ momentum smearing~\cite{Bali:2016lva} for the quark propagators, one-step hypercubic (HYP) smearing~\cite{Hasenfratz:2001hp,DeGrand:2002vu} for the spatial gauge links in the nonlocal operators, and a kinematically enhanced source-side interpolator~\cite{Zhang:2025hyo,Reitinger:2026hta} designed to increase the overlap with the boosted ground state.

%%%%%%%%%%%%%%%%%%%%%%%%%%%%%%%%%%%%%%%%%%%%%%%%%%%%%%%%%%%%%%%%%%%%
\begin{table}[htbp]
    \centering
    \caption{Lattice ensembles used in this work.}
    \label{tab:Ensembles}
    \renewcommand{\arraystretch}{1.15}
    \setlength{\tabcolsep}{3pt}
    \begin{tabular}{c c c c c}
        \hline\hline
        Ensemble & $a({\rm fm})$ & $m_\pi({\rm MeV})$ & $P^z({\rm GeV})$ & $n_{\rm stat}$ \\
        \hline
        C24P29 & $0.1052$ & $292.3$ & $1.96, 2.45, 2.94$ & $864\times36$ \\
        C32P23 & $0.1052$ & $227.9$ & $1.84, 2.21, 2.57, 2.94$ & $954\times32$ \\
        C48P14 & $0.1052$ & $136.4$ & $1.96, 2.45, 2.94$ & $302\times64$ \\
        F32P30 & $0.0775$ & $300.4$ & $2.00, 2.49, 2.99$ & $777\times32$ \\
        F32P21 & $0.0775$ & $210.3$ & $2.00, 2.49, 2.99$ & $459\times64$ \\
        G36P29 & $0.0689$ & $297.2$ & $2.00, 2.50, 3.00$ & $656\times32$ \\
        H48P32 & $0.0520$ & $316.6$ & $1.98, 2.48, 2.98$ & $550\times54$ \\
        \hline\hline
    \end{tabular}
\end{table}
%%%%%%%%%%%%%%%%%%%%%%%%%%%%%%%%%%%%%%%%%%%%%%%%%%%%%%%%%%%%%%%%%%%%

For each ensemble and each available momentum, the coordinate-space quasi-DAs are extracted from corresponding two-point correlation functions.
After hybrid renormalization, large-$\lambda$ extrapolation, two-dimensional Fourier transform, and LaMET matching, we obtain the matched LCDAs $\phi(x_1,x_2)|_{a,m_\pi,P^z}$ at finite lattice spacings, unphysical pion masses, and finite hadron momenta. 
The residual dependence on $a$, $m_\pi$, and $P^z$ is then removed by a physical-limit extrapolation performed independently at each momentum-fraction point $(x_1,x_2)$.

The combined extrapolation ansatz used to obtain the central value of result in our analysis is
\begin{equation}\label{eq:joint_extrap}
\begin{aligned}
    \phi(x_1,x_2)&|_{a,m_\pi,P^z} = \phi_{\rm phys}(x_1,x_2) \\
    & + \frac{A(x_1,x_2)}{(P^z)^2} + \left( m_\pi^2 - m_{\pi,\rm phys}^2 \right) B(x_1,x_2)\\
    & + a^2 \left[ D_1(x_1,x_2) + (P^z)^2 D_2(x_1,x_2) \right]\ .
\end{aligned}
\end{equation}
Here $\phi_{\rm phys}(x_1,x_2)$ denotes the desired LCDA in the continuum, physical-pion-mass, and infinite-momentum limits.
The term proportional to $1/(P^z)^2$ parametrizes the leading LaMET power correction, while the term proportional to $(m_\pi^2-m_{\pi,\rm phys}^2)$ describes the leading pion-mass dependence.
The two terms proportional to $a^2$ account for the momentum-independent discretization effect $a^2 D_1$, and momentum-dependent artifact $a^2(P^z)^2D_2$ associated with boosted hadron states.

The same analysis and extrapolation strategies are applied to all leading-twist $V$, $A$, and $T$ amplitudes. 
The statistical uncertainty is propagated through the full procedure using jackknife samples.
The systematic uncertainties include the residual dependence on the renormalization and matching scale $\mu$, the large-$\lambda$ extrapolation, and the physical-limit extrapolation in $a$, $m_\pi$, and $P^z$.

%%%%%%%%%%%%%%%%%%%%%%%%%%%%%%%%%%%%%%%%%%%%%%%%%%%%%%%%%%%%%%%%%%%%
\begin{figure}[htbp]
    \centering
    \includegraphics[width=0.95\linewidth]{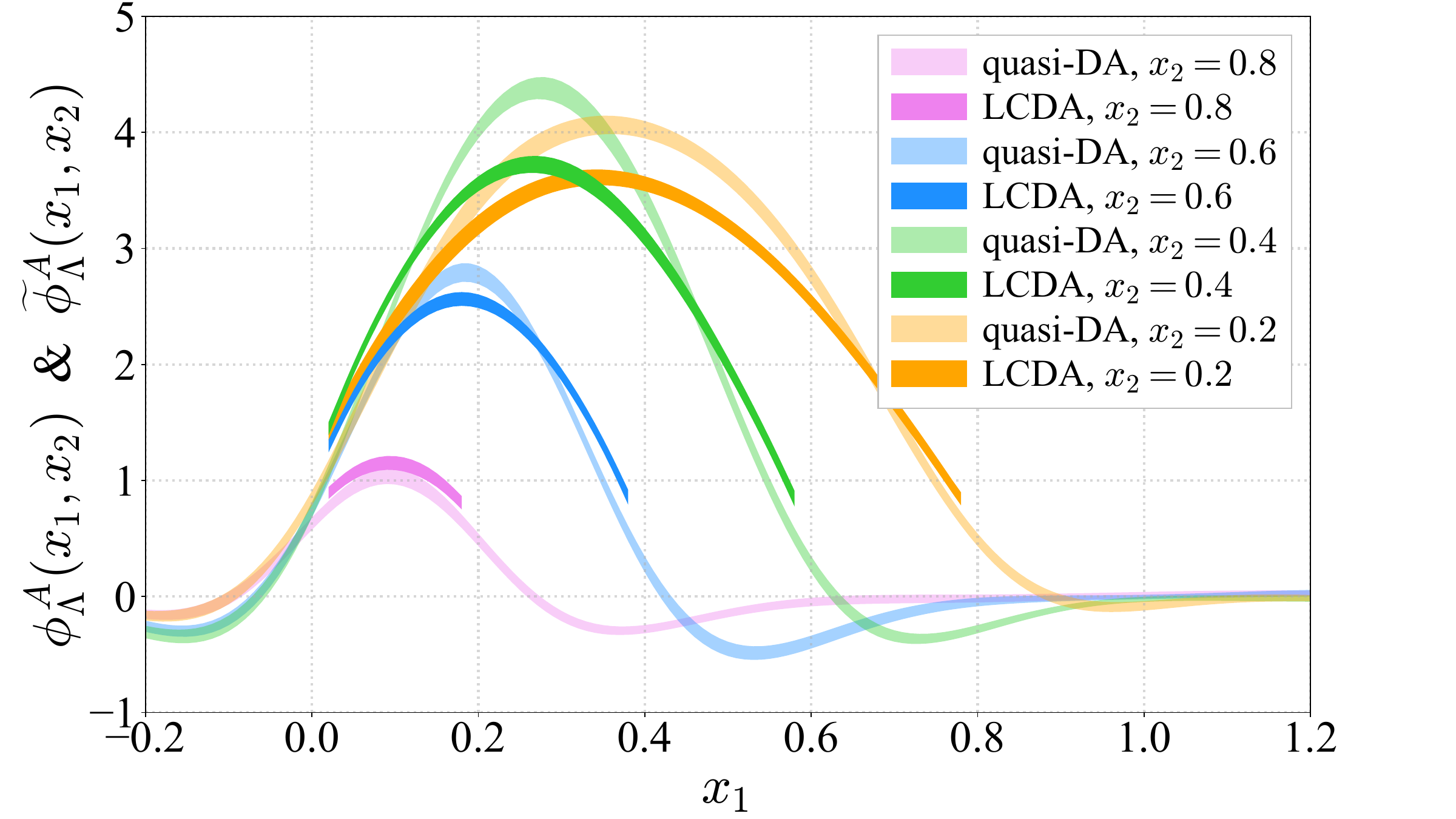}
    \caption{Comparison of the hybrid-renormalized quasi-DA $\widetilde\phi_\Lambda^A(x_1,x_2)$ and the LCDA $\phi_\Lambda^A(x_1,x_2)$ after one-loop LaMET matching, shown on the H48P32 ensemble at $P^z\approx 3~{\rm GeV}$ for several fixed values of $x_2$.}
    \label{fig:matching}
\end{figure}
%%%%%%%%%%%%%%%%%%%%%%%%%%%%%%%%%%%%%%%%%%%%%%%%%%%%%%%%%%%%%%%%%%%%

%%%%%%%%%%%%%%%%%%%%%%%%%%%%%%%%%%%%%%%%%%%%%%%%%%%%%%%%%%%%%%%%%%%%
\begin{figure}[htbp]
    \centering
    \includegraphics[width=0.95\linewidth]{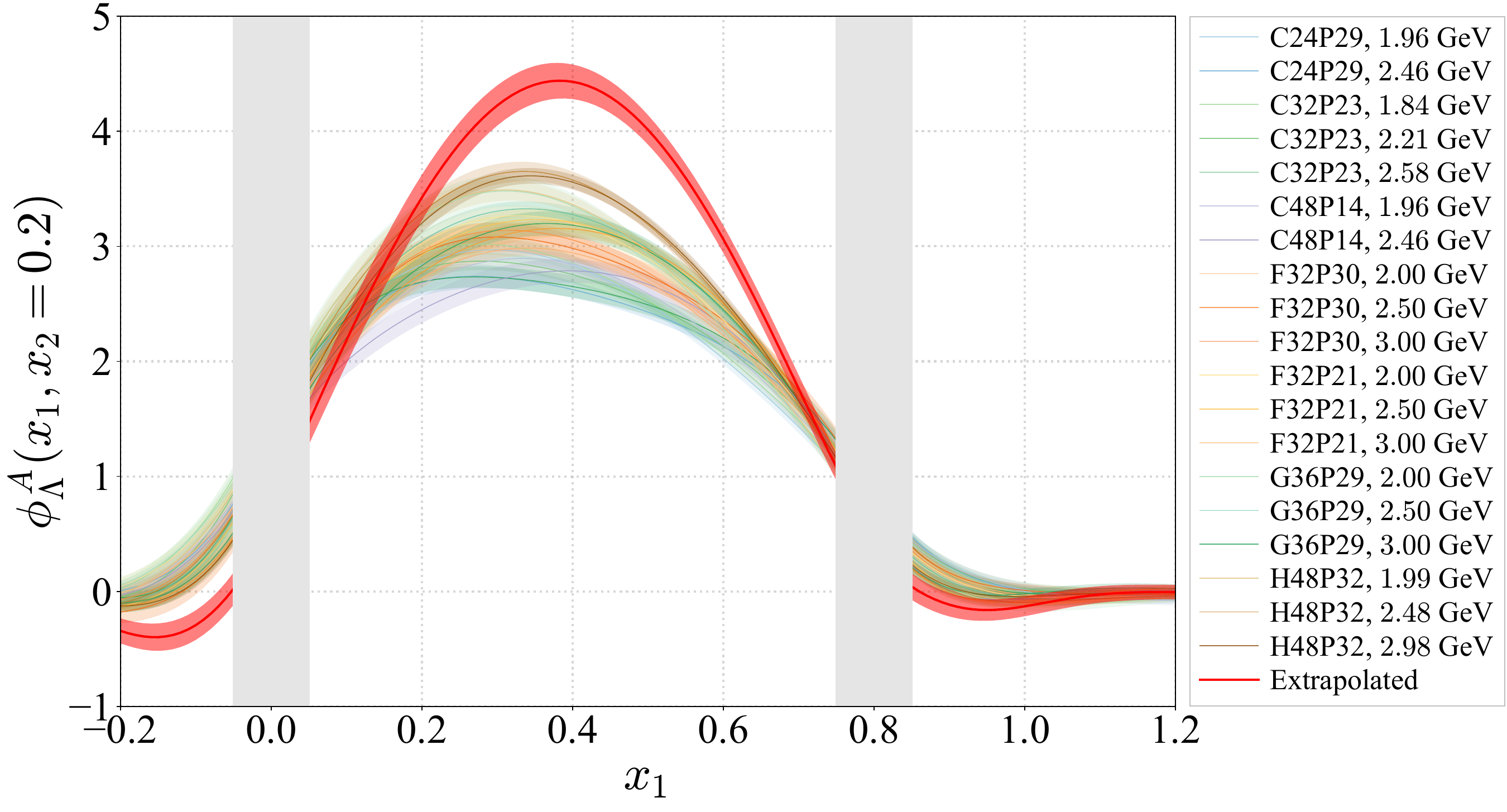}
    \caption{Combined extrapolation of the $\Lambda$ $A$ LCDA at fixed $x_2=0.2$. The colored curves show the matched LCDAs obtained from different ensembles and hadron momenta, while the red curve denotes the extrapolated result in the continuum, physical-pion-mass, and infinite-momentum limits. The gray bands indicate endpoint regions where LaMET power corrections are enhanced.}
    \label{fig:joint_extrap}
\end{figure}
%%%%%%%%%%%%%%%%%%%%%%%%%%%%%%%%%%%%%%%%%%%%%%%%%%%%%%%%%%%%%%%%%%%%

%%%%%%%%%%%%%%%%%%%%%%%%%%%%%%%%%%%%%%%%%%%%%%%%%%%%%%%%%%%%%%%%%%%%
%%%%%%%%%%%%%%%%%%%%%%%%%%%%%%%%%%%%%%%%%%%%%%%%%%%%%%%%%%%%%%%%%%%%
\paragraph{Results for leading-twist $\Lambda$ LCDAs.} We now present the numerical results for the leading-twist $\Lambda$-baryon LCDAs.
We first illustrate the impact of LaMET matching on the quasi-DAs before proceeding to the physical-limit extrapolation.
As a representative example, Fig.~\ref{fig:matching} compares the hybrid-renormalized quasi-DAs $\widetilde{\phi}^A_\Lambda$ with the corresponding LCDAs $\phi^A_\Lambda$ after one-loop LaMET matching on the H48P32 ensemble at $a\approx0.052~{\rm fm}$ and $P^z\approx 3~{\rm GeV}$.
Since the baryon LCDAs depend on two independent momentum fractions, the comparison is shown by one-dimensional slices at several fixed $x_2$.
The matching produces a visible momentum-fraction-dependent modification of the distribution, rather than a simple overall rescaling.
This demonstrates the importance of matching in converting the finite-momentum quasi-DAs into the LCDAs.

%%%%%%%%%%%%%%%%%%%%%%%%%%%%%%%%%%%%%%%%%%%%%%%%%%%%%%%%%%%%%%%%%%%%
\begin{figure*}[htbp]
    \centering
    \subfloat{
        \centering
        \includegraphics[width=0.32\textwidth]{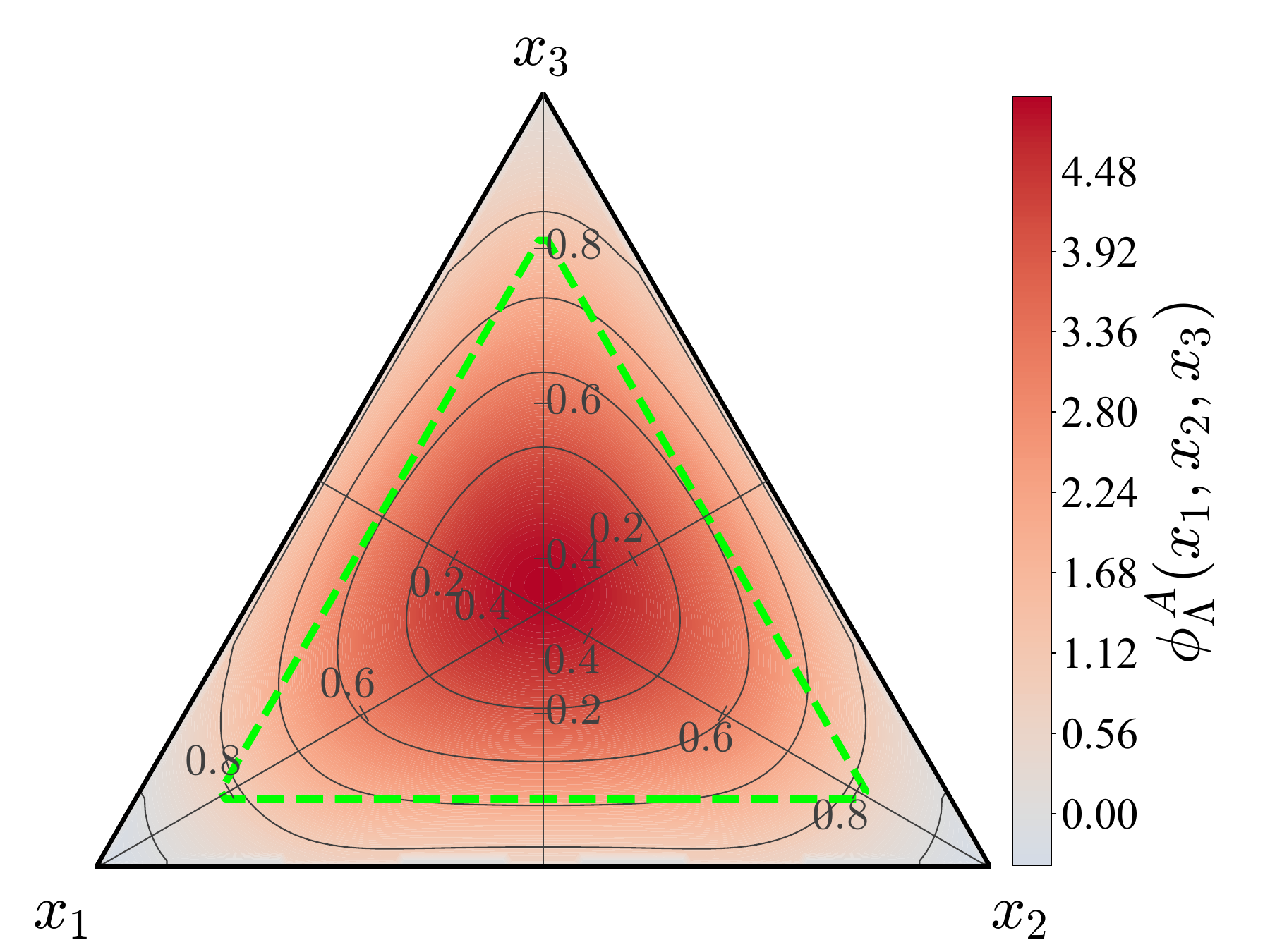}
        }
    \subfloat{
        \centering
        \includegraphics[width=0.32\textwidth]{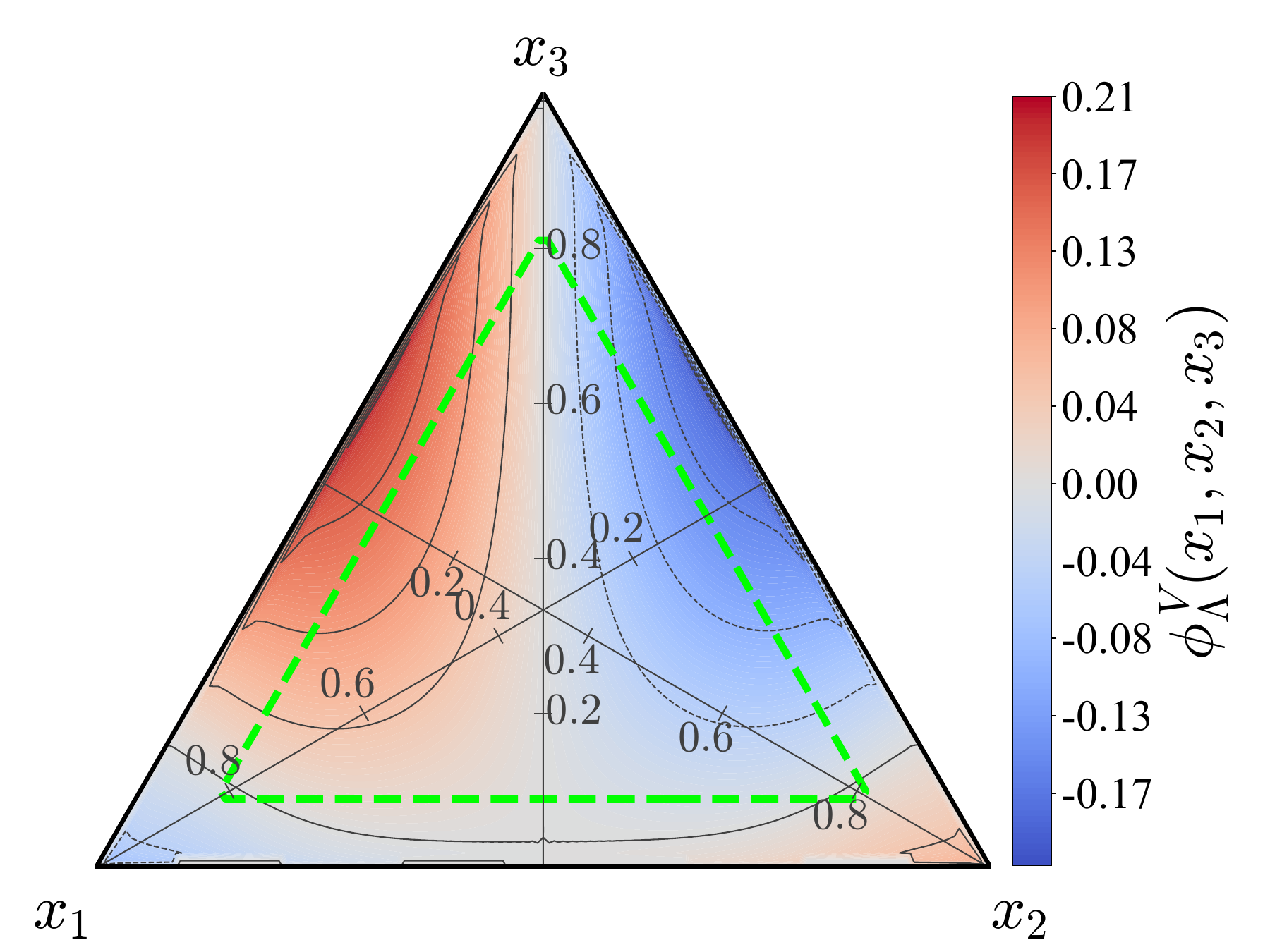}
        }
    \subfloat{
        \centering
        \includegraphics[width=0.32\textwidth]{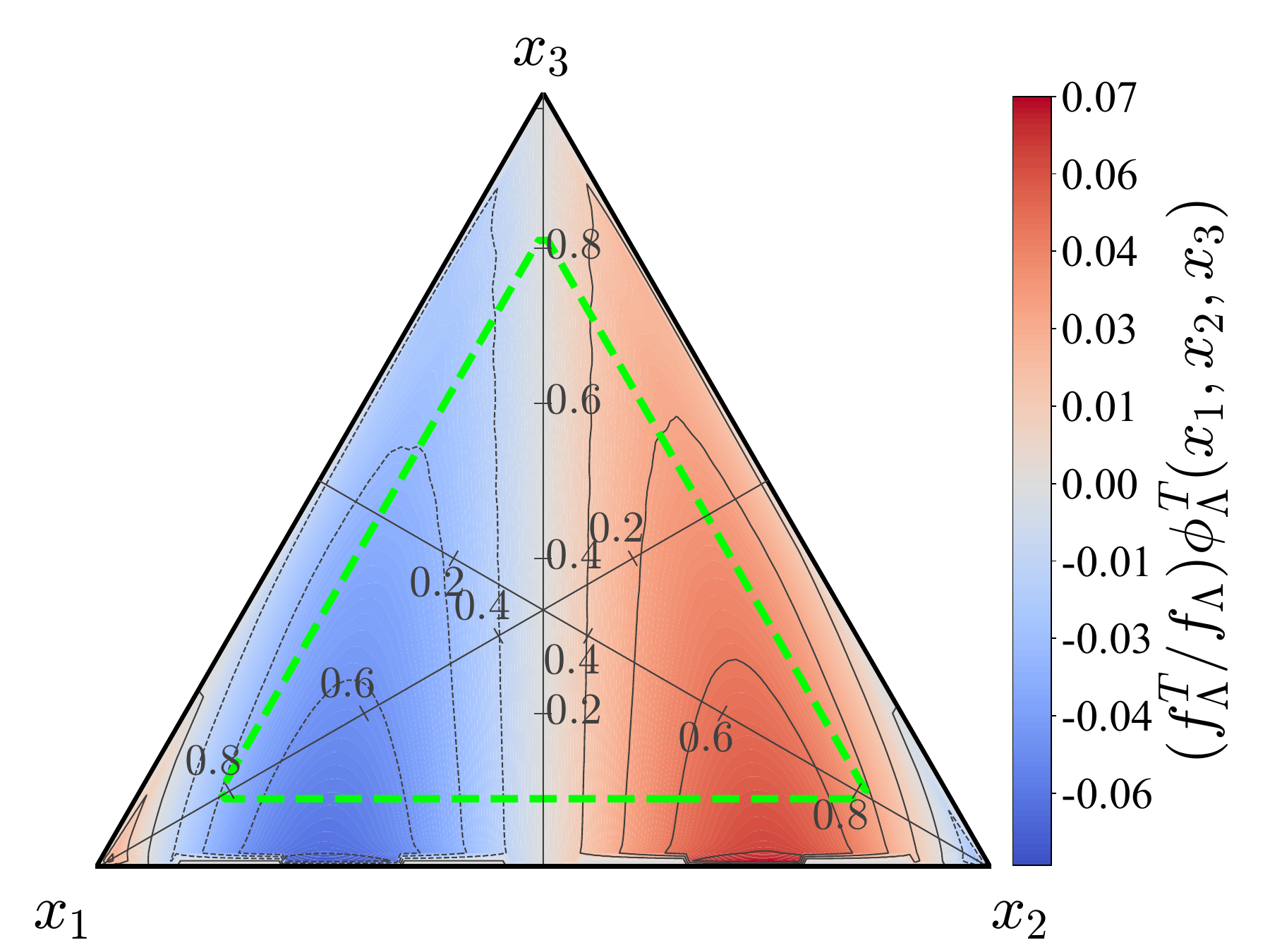}
        }
        \\
    \subfloat{
        \centering
        \includegraphics[width=0.32\textwidth]{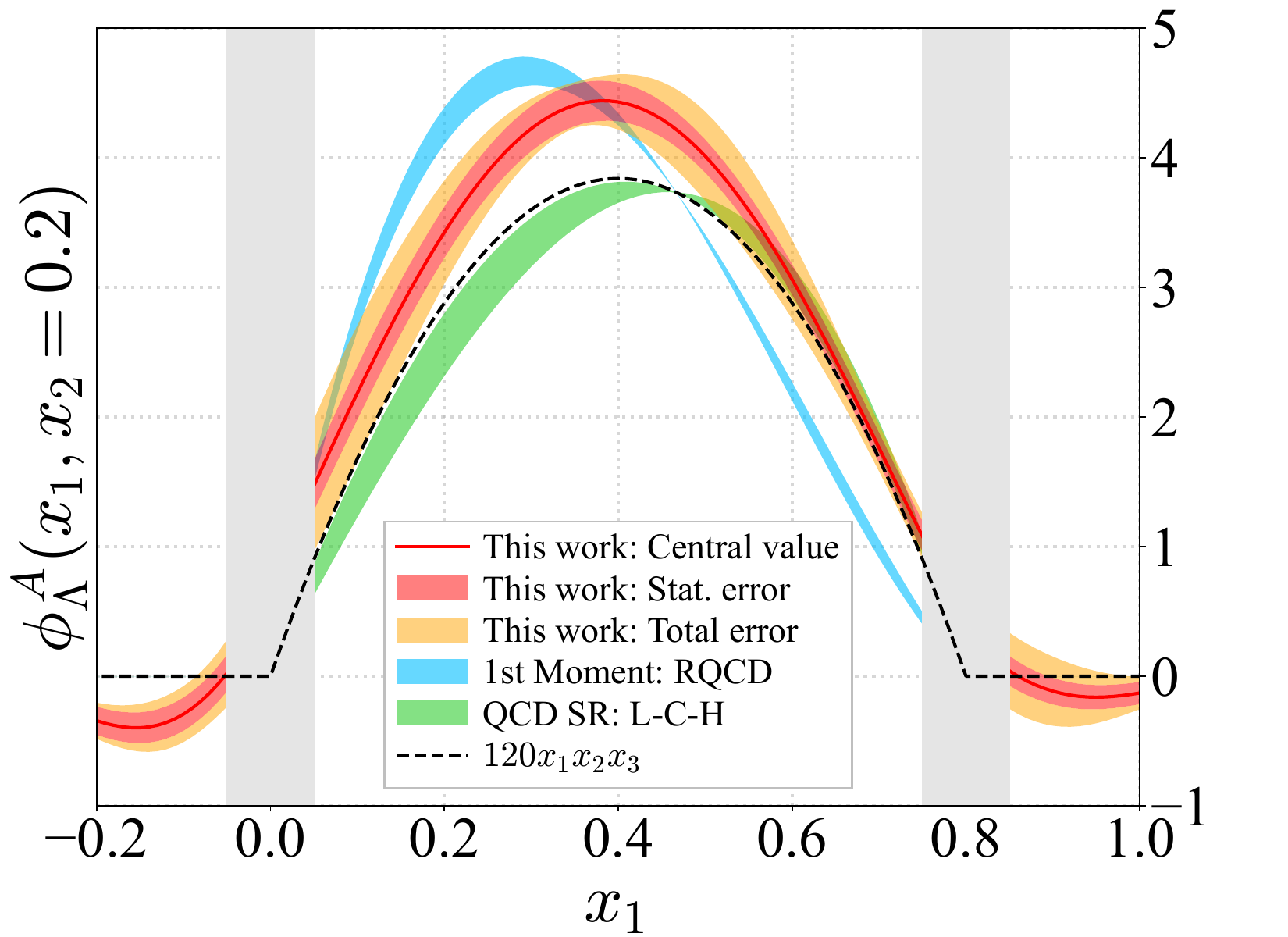}
        }
    \subfloat{
        \centering
        \includegraphics[width=0.32\textwidth]{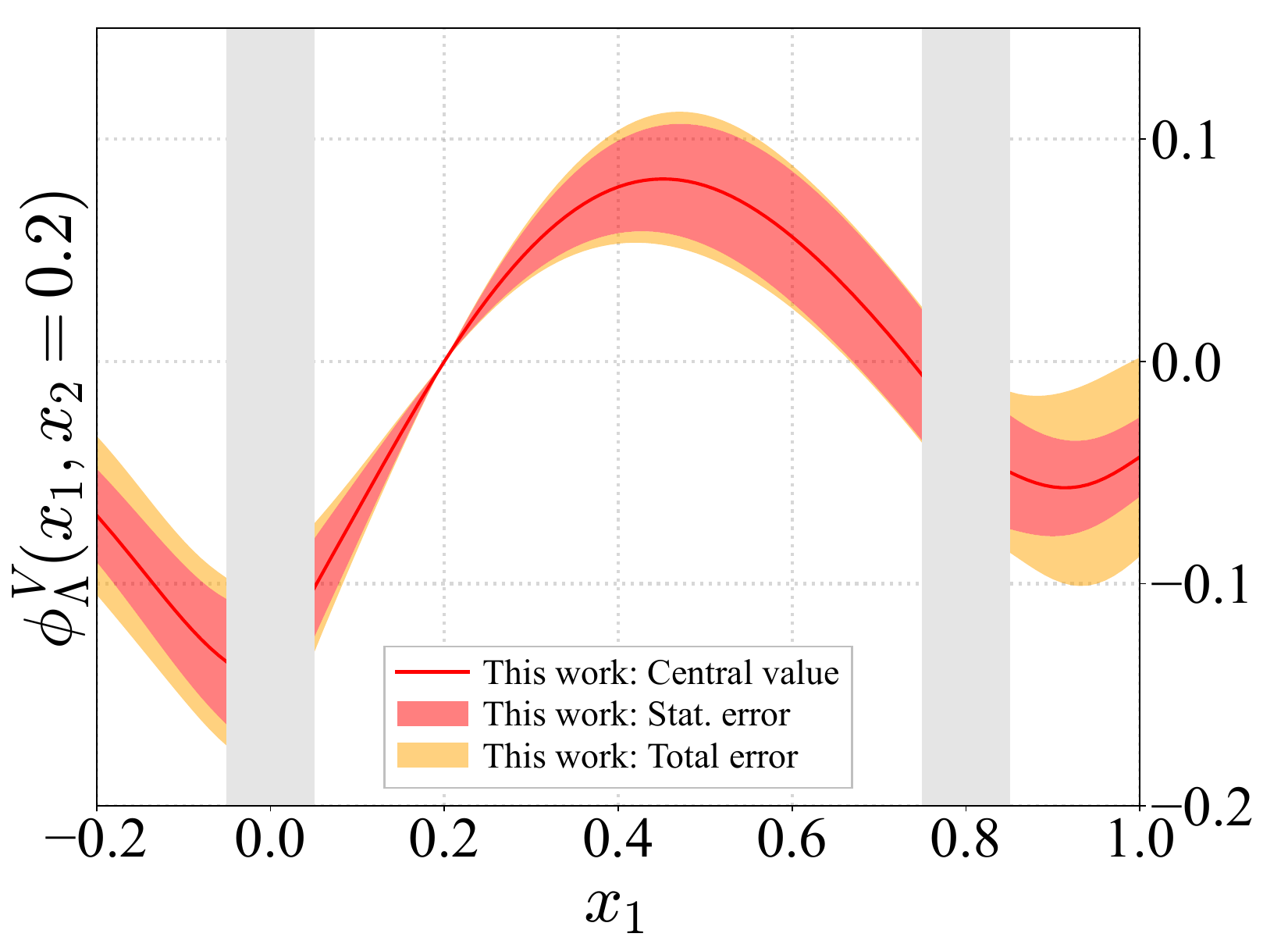}
        }
    \subfloat{
        \centering
        \includegraphics[width=0.32\textwidth]{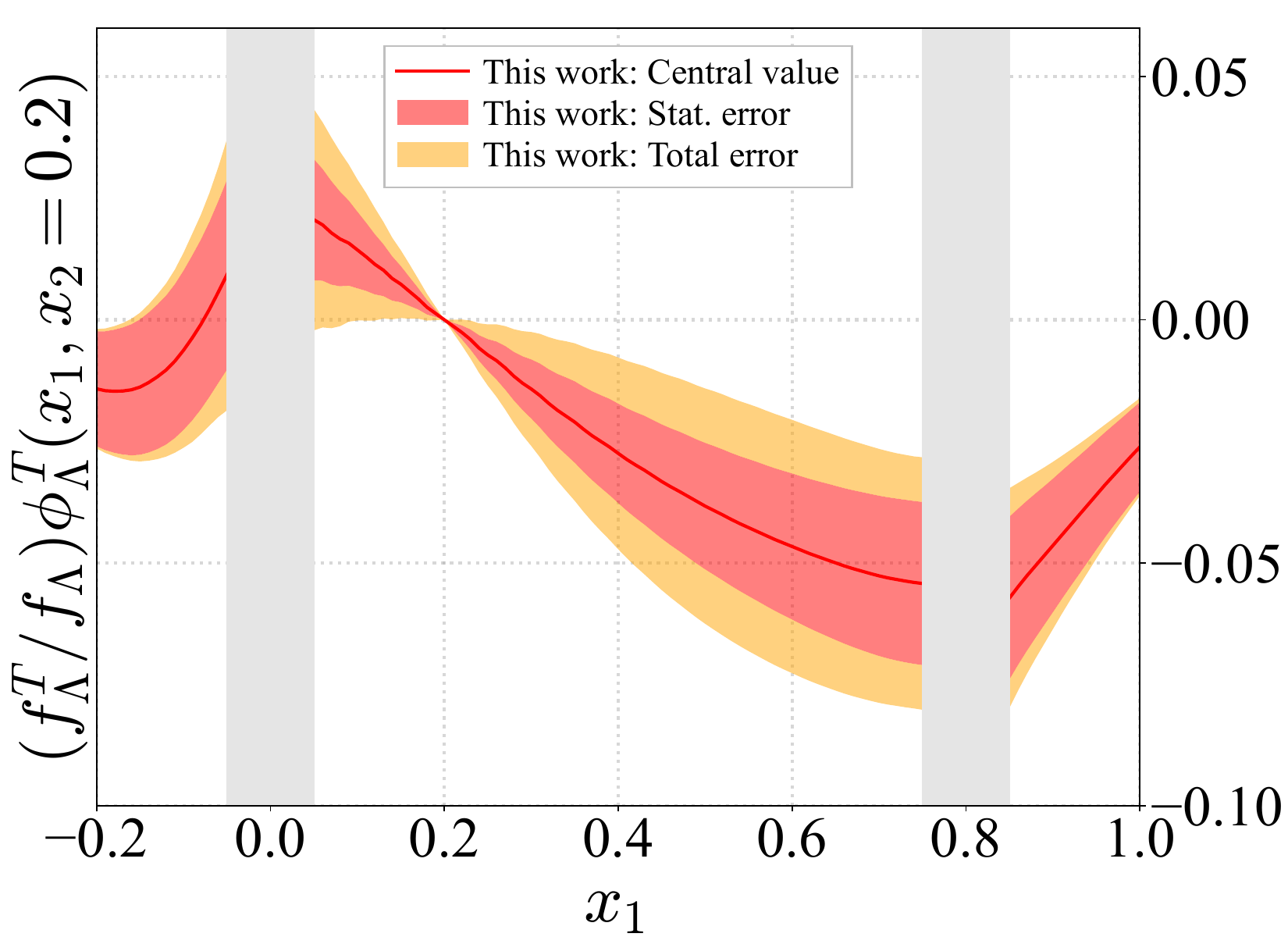}
        }
    \caption{Final results for the leading-twist $\Lambda$ LCDAs. The upper row shows the two-dimensional distributions of the $A$, $V$, and $T$ amplitudes in the physical triangular region $0\le x_1,x_2,x_3\le 1$. The green dashed line indicates the LaMET reliable region, $x_{1,2,3}\ge 0.1$, where high-power corrections are expected to be under better control. The lower row shows the corresponding one-dimensional cuts at fixed $x_2=0.2$, with statistical and total uncertainties. For the $A$ amplitude, we also compare with the asymptotic form $120x_1x_2x_3$, the lattice-OPE first-moment reconstruction~\cite{Bali:2024oxg}, and QCD sum-rule model~\cite{Liu:2014uha}.}
    \label{fig:final_results}
\end{figure*}
%%%%%%%%%%%%%%%%%%%%%%%%%%%%%%%%%%%%%%%%%%%%%%%%%%%%%%%%%%%%%%%%%%%%

% After matching, the LCDAs obtained on individual ensembles still contain residual dependence on the lattice spacing, the pion mass, and the finite hadron momentum.

After matching, we remove the residual dependence on the $a$, $m_\pi$, and finite $P^z$ in the LCDAs obtained on individual ensembles, using the combined extrapolation in Eq.~(\ref{eq:joint_extrap}).
Fig~\ref{fig:joint_extrap} shows this procedure for the $\Lambda$ $A$ amplitude at the representative slice $x_2=0.2$.
The colored bands denote the matched LCDAs from different ensembles and momenta, whose variations are described by the finite-$a$, unphysical-$m_\pi$, and finite-$P^z$ terms.
The red band represents the resulting LCDA after extrapolation to the continuum, physical-pion-mass, and infinite-momentum limits.
The separation between the extrapolated curve and the finite-momentum data indicates that residual LaMET power corrections remain numerically relevant for baryon observables at the currently accessible momenta.
The gray bands indicate endpoint regions, where the power corrections in Eq.~(\ref{eq:LaMET}) are expected to be enhanced.

The final results for all three leading-twist structures are shown in Fig.~\ref{fig:final_results}.
The upper panels display the full two-dimensional distributions in the physical momentum-fraction region, while the lower panels show representative slice at fixed $x_2=0.2$ with statistical and total uncertainties.
The $\Lambda$ $A$ amplitude is symmetric under $x_1\leftrightarrow x_2$ and gives the dominant contribution.
It exhibits a broad peak in the interior of the physical triangle region, with a mild displacement toward the $x_3$ direction.
This indicates that the strange quark carries a slightly larger share of the longitudinal momentum than the two light quarks.
For the $A$ amplitude, we also compare with the asymptotic form $120x_1x_2x_3$, the lattice-OPE first-moment reconstruction~\cite{Bali:2024oxg} , and QCD sum-rule model~\cite{Liu:2014uha}.
On the $x_2=0.2$ slice, our result differs visibly from several of these previously used inputs, highlighting the importance of a direct $x$-dependent lattice determination beyond low-moment constraints.

The $V$ and $T$ amplitudes are antisymmetric under $x_1\leftrightarrow x_2$, change sign across the $x_1=x_2$ line, and vanish on that line as required by the $\Lambda$ exchange symmetries.
Their nonzero structures away from the symmetric line are clearly resolved, although their magnitudes are smaller than that of the dominant $A$ amplitude.
This provides direct $x$-dependent information on the antisymmetric amplitudes beyond the moment constraints available from local operators.
Together, the $V$, $A$, and $T$ results provide a complete first-principles determination of the leading-twist $\Lambda$-baryon LCDAs and direct lattice-QCD access to multi-dimensional light-cone structure in momentum space.

%%%%%%%%%%%%%%%%%%%%%%%%%%%%%%%%%%%%%%%%%%%%%%%%%%%%%%%%%%%%%%%%%%%%
%%%%%%%%%%%%%%%%%%%%%%%%%%%%%%%%%%%%%%%%%%%%%%%%%%%%%%%%%%%%%%%%%%%%
\paragraph{Phenomenological implications and discussion.} To facilitate phenomenological applications, we further provide a compact parameterization of the lattice-determined $\Lambda$ LCDAs.
Following the conventions of Refs.~\cite{Braun:2008ia,Huang:2025bbk}, we project our results onto the corresponding conformal polynomial basis in the LaMET reliable region and determine the expansion coefficients for the $\Lambda$-baryon $V$, $A$, and $T$ amplitudes.
The recommended second-order parameterization is summarized in Table~\ref{tab:lambda_param_fit}. This truncation captures the main features resolved by the lattice data while avoiding unnecessary higher-order flexibility. Stability checks with first- and third-order fits are given in the Supplemental Material~\cite{supplemental}.
Since the fit is performed globally over the two-dimensional $(x_1,x_2)$ momentum-fraction plane, the finite set of coefficients is constrained by the full lattice-determined shape rather than by a small number of local moments, leading to relatively small statistical uncertainties in the fitted results.

%%%%%%%%%%%%%%%%%%%%%%%%%%%%%%%%%%%%%%%%%%%%%%%%%%%%%%%%%%%%%%%%%%%%
\begin{table*}[htbp]
    \centering
    \caption{Recommended second-order parameterization of the lattice-determined $\Lambda$ LCDAs. The coefficients $\phi_{\Lambda,nm}$ determine the $A$ and $V$ amplitudes, while $\widetilde{\pi}_{\Lambda,nm}$ determine the $T$ amplitude. To suppress the impact of endpoint-enhanced higher-power corrections, the fit is restricted to the region $x_{1,2,3}\geq 0.1$.}
    \label{tab:lambda_param_fit}
    \renewcommand{\arraystretch}{1.15}
    \setlength{\tabcolsep}{8pt}
    \begin{tabular}{c c c c c c c}
         \hline\hline
         $\phi_{\Lambda,10}$ & $\phi_{\Lambda,11}$ & $\phi_{\Lambda,20}$ & $\phi_{\Lambda,21}$ & $\phi_{\Lambda,22}$ & $\widetilde{\pi}_{\Lambda,10}$ & $\widetilde{\pi}_{\Lambda,21}$ \\
         \hline
         $0.0343(12)$ & $0.00543(40)$ & $0.0012(14)$ & $-0.0267(18)$ & $-0.0087(14)$ &$0.00750(10)$ & $0.00043(14)$ \\
         \hline\hline
    \end{tabular}
\end{table*}
%%%%%%%%%%%%%%%%%%%%%%%%%%%%%%%%%%%%%%%%%%%%%%%%%%%%%%%%%%%%%%%%%%%%

%%%%%%%%%%%%%%%%%%%%%%%%%%%%%%%%%%%%%%%%%%%%%%%%%%%%%%%%%%%%%%%%%%%%
\begin{table}[htbp]
    \centering
    \caption{Impacts on $\Lambda$ form factor $F_1^\Lambda$ at $Q^2=20~{\rm GeV}^2$.}
    \label{tab:lambda_ff_f1}
    \renewcommand{\arraystretch}{1.15}
    \setlength{\tabcolsep}{6pt}
    \begin{tabular}{c c c c c}
        \hline\hline
        $F_1^\Lambda$ & Asymptotic & This work & Impact \\
        \hline
        LO  & $0.00425$ & $0.00371$ & $12.7\%$\\
        NLO & $0.00913$ & $0.00806$ & $11.7\%$\\
        \hline\hline
    \end{tabular}
\end{table}
%%%%%%%%%%%%%%%%%%%%%%%%%%%%%%%%%%%%%%%%%%%%%%%%%%%%%%%%%%%%%%%%%%%%

As a representative test of the phenomenological impact from the lattice-determined LCDAs, we evaluate the $\Lambda$ electromagnetic form factor $F_1^\Lambda$ following the perturbative-QCD hard-scattering formula of Ref.~\cite{Chernyak:1987nu}, 
supplemented by the corresponding NLO contribution for $\Lambda$ obtained from an extension of Ref.~\cite{Chen:2024fhj}.
The comparison with the commonly used asymptotic form is shown in Table~\ref{tab:lambda_ff_f1}.
At $Q^2=20~{\rm GeV}^2$, replacing the asymptotic form by our lattice-determined LCDAs changes $F_1^\Lambda$ from $0.0043$ to $0.0037$ at LO, corresponding to a $12.7\%$ effect.
At NLO, this change is from $0.0091$ to $0.0081$, giving an $11.7\%$ effect.
This visible $\mathcal{O}(10 \%)$ shift supports the observation from recent NLO studies, once the perturbative hard kernel is under control, the baryon LCDAs input can become the dominant phenomenological ambiguity in hard-scattering predictions.
Our result therefore gives a quantitative illustration of why the full two-dimensional LCDAs, rather than only their asymptotic forms or lowest moments, are needed for precision baryonic phenomenology.

%\Vio{ww: Ref.[35] is the calculation for proton, but here is $\Lambda$. Compared to other uncertainties how important is the $10\%$ correction? }

%%%%%%%%%%%%%%%%%%%%%%%%%%%%%%%%%%%%%%%%%%%%%%%%%%%%%%%%%%%%%%%%%%%%
%%%%%%%%%%%%%%%%%%%%%%%%%%%%%%%%%%%%%%%%%%%%%%%%%%%%%%%%%%%%%%%%%%%%
\paragraph{Summary.} We have presented the first complete first-principles determination of the leading-twist $\Lambda$-baryon LCDAs.
Combining boosted nonlocal three-quark matrix elements with hybrid renormalization, large-$\lambda$ extrapolation, perturbative matching, and controlled extrapolation to continuum, physical-pion-mass, and infinite-momentum limits, we obtain all $V$, $A$, and $T$ amplitudes at physical point as two-dimensional functions of the valence-quark momentum fractions.
These results provide a direct lattice-QCD access of the intrinsic multi-dimensional light-cone structures to baryons, resolving both the dominant symmetric amplitude and the antisymmetric structures beyond the local-moment constraints.
For phenomenological use, we provide compact parameterizations and find that the lattice-determined shape induces an $\mathcal{O}(10 \%)$ shift in a representative $\Lambda$ electromagnetic form factor compare with the asymptotic form.
This work establishes a first-principles benchmark for multi-dimensional baryon light-cone structure and for quantitative applications to baryonic hard exclusive and CP-sensitive processes.\\

%%%%%%%%%%%%%%%%%%%%%%%%%%%%%%%%%%%%%%%%%%%%%%%%%%%%%%%%%%%%%%%%%%%%
%%%%%%%%%%%%%%%%%%%%%%%%%%%%%%%%%%%%%%%%%%%%%%%%%%%%%%%%%%%%%%%%%%%%
%\paragraph{Acknowledgment.} 
\begin{acknowledgments}
\paragraph{Acknowledgments.} We thank Long-Bin Chen and Feng Feng for providing the code of NLO $\Lambda$ electromagnetic form factor as an extension of Ref.~\cite{Chen:2024fhj}, and we thank Jia-Jie Han, Yushan Su for valuable discussions.
We thank the CLQCD collaboration for providing us the gauge configurations with dynamical fermions~\cite{CLQCD:2023sdb}, which are generated on the HPC Cluster of ITP-CAS, the Southern Nuclear Science Computing Center (SNSC), the Siyuan-1 cluster supported by the Center for High Performance Computing at Shanghai Jiao Tong University and the Dongjiang Yuan Intelligent Computing Center.
This work is supported in part by Natural Science Foundation of China under grant No.12575084, 12375069, 12575085, 12222503, 12293060, 12293062, 12435002, 12293065, 12047503, 12125503, 12305103, 12375080, 12275277 and 12435004. M.-H.~C. is supported by the National Science Centre (Poland) grant OPUS No.2021/43/B/ST2/00497. J.~H is supported by Natural Science Foundation of Guangdong Province under Grant No.2025A1515012199. Y.-B.~Y is supported by National Key R\&D Program of China No.2024YFE0109800 and Strategic Priority Research Program of Chinese Academy of Sciences, Grant No.YSBR-101. J.-H.~Z is supported by the Ministry of Science and Technology of China under Grant No.2024YFA1611004, and CUHK-Shenzhen under grant No.UDF01002851.
The computations in this work were run on the Siyuan-1 cluster supported by the Center for High Performance Computing at Shanghai Jiao Tong University, Southern Nuclear Science Computing Center (SNSC), and Advanced Computing East China Sub-center. The LQCD simulations were performed using the PyQUDA software suite~\cite{Jiang:2024lto} and QUDA~\cite{Clark:2009wm,Babich:2011np,Clark:2016rdz} through HIP programming model~\cite{Bi:2020wpt}. 
\end{acknowledgments}

%%%%%%%%%%%%%%%%%%%%%%%%%%%%%%%%%%%%%%%%%%%%%%%%%%%%%%%%%%%%%%%%%%%%
%%%%%%%%%%%%%%%%%%%%%%%%%%%%%%%%%%%%%%%%%%%%%%%%%%%%%%%%%%%%%%%%%%%%
% \bibliographystyle{apsrev4-2}
\bibliography{ref}

%%%%%%%%%%%%%%%%%%%%%%%%%%%%%%%%%%%%%%%%%%%%%%%%%%%%%%%%%%%%%%%%%%%%
% This paragraph should be kept for the arXiv submission
% and commented out for the PRL submission.
\clearpage
\onecolumngrid
\def\SupplementInMain{}
\ifdefined\SupplementInMain
  % Included in the main arXiv file: do not load document class or packages.
  \section*{Supplemental Material}
  
\else
  \documentclass[aps,prl,onecolumn,10pt,preprintnumbers,nofootinbib,superscriptaddress,floatfix]{revtex4-2}

  \usepackage{amsmath}
  \usepackage{amssymb}
  \usepackage{graphicx}
  \usepackage{mathrsfs}
  \usepackage{slashed}
  \usepackage{array}
  \usepackage[dvipsnames]{xcolor}
  \usepackage{indentfirst}
  \usepackage{multirow}
  \usepackage[caption=false]{subfig}
  \usepackage{placeins}
  
  \usepackage{xr-hyper}
  \usepackage[colorlinks=true, allcolors=blue]{hyperref}
  \usepackage{cleveref}
  
  \externaldocument{Lettermain}
  
  \newcommand{\rme}{\mathrm{e}}
  \newcommand{\rmi}{\mathrm{i}}
  \newcommand{\rmd}{\mathrm{d}}

  \begin{document}

  \title{Supplemental Material for:\\Complete Access to Leading-Twist \texorpdfstring{$\Lambda$}{Lambda}-Baryon Light-Cone Distribution Amplitudes from Lattice QCD}

  \collaboration{\bf{Lattice Parton Collaboration ($\rm {\bf LPC}$)}}
  
  \author{\includegraphics[scale=0.10]{logo.png}\\Mu-Hua Zhang}
  \affiliation{State Key Laboratory of Dark Matter Physics, Key Laboratory for Particle Astrophysics and Cosmology (MOE), Shanghai Key Laboratory for Particle Physics and Cosmology, Tsung-Dao Lee Institute and School of Physics and Astronomy, Shanghai Jiao Tong University, Shanghai 200240, China}
  
  \author{Haoyang Bai}
  \affiliation{Institute of High Energy Physics, CAS, Beijing 100049, China}
  \affiliation{School of Physics, University of Chinese Academy of Sciences, Beijing 100049, China}
  
  \author{Min-Huan Chu}
  \affiliation{Faculty of Physics and Astronomy, Adam Mickiewicz University, ul.\ Uniwersytetu Pozna\'nskiego 2, 61-614 Pozna\'{n}, Poland}
  
  \author{Jun Hua}
  \thanks{Corresponding author.}
  \email{junhua@scnu.edu.cn}
  \affiliation{State Key Laboratory of Nuclear Physics and Technology, Institute of Quantum Matter, South China Normal University, Guangzhou 510006, China}
  \affiliation{Guangdong Basic Research Center of Excellence for Structure and Fundamental Interactions of Matter, Guangdong Provincial Key Laboratory of Nuclear Science, Guangzhou 510006, China}
  
  \author{Xiangdong Ji}
  \affiliation{Tsung-Dao Lee Institute and School of Physics and Astronomy,
  Shanghai Jiao Tong University, Shanghai 201210, China}
  
  \author{Xiangyu Jiang}
  \affiliation{CAS Key Laboratory of Theoretical Physics, Institute of Theoretical Physics, Chinese Academy of Sciences, Beijing 100190, China}
  
  \author{Jian Liang}
  \affiliation{State Key Laboratory of Nuclear Physics and Technology, Institute of Quantum Matter, South China Normal University, Guangzhou 510006, China}
  \affiliation{Guangdong Basic Research Center of Excellence for Structure and Fundamental Interactions of Matter, Guangdong Provincial Key Laboratory of Nuclear Science, Guangzhou 510006, China}

  \author{Cai-Dian L\"u}
  \affiliation{Institute of High Energy Physics, CAS, Beijing 100049, China}
  \affiliation{School of Physics, University of Chinese Academy of Sciences, Beijing 100049, China}
  
  \author{Andreas Sch\"afer}
  \affiliation{Institut f\"ur Theoretische Physik, Universit\"at Regensburg, D-93040 Regensburg, Germany}
  
  \author{Wei Wang}
  \thanks{Corresponding author.}
  \email{wei.wang@sjtu.edu.cn}
  \affiliation{State Key Laboratory of Dark Matter Physics, Key Laboratory for Particle Astrophysics and Cosmology (MOE), Shanghai Key Laboratory for Particle Physics and Cosmology, School of Physics and Astronomy, Shanghai Jiao Tong University, Shanghai 200240, China}
  \affiliation{Southern Center for Nuclear-Science Theory (SCNT), Institute of Modern Physics, Chinese Academy of Sciences, Huizhou 516000, Guangdong Province, China}
  
  \author{Yi-Bo Yang}
  \affiliation{CAS Key Laboratory of Theoretical Physics, Institute of Theoretical Physics, Chinese Academy of Sciences, Beijing 100190, China}
  \affiliation{School of Fundamental Physics and Mathematical Sciences, Hangzhou Institute for Advanced Study, UCAS, Hangzhou 310024, China}
  \affiliation{International Centre for Theoretical Physics Asia-Pacific, Beijing/Hangzhou, China}
  \affiliation{School of Physical Sciences, University of Chinese Academy of Sciences,
  Beijing 100049, China}
  
  \author{Jian-Hui Zhang}
  \affiliation{School of Science and Engineering, The Chinese University of Hong Kong, Shenzhen 518172, China}
  
  \author{Jia-Lu Zhang}
  \affiliation{State Key Laboratory of Dark Matter Physics, Key Laboratory for Particle Astrophysics and Cosmology (MOE), Shanghai Key Laboratory for Particle Physics and Cosmology, Tsung-Dao Lee Institute and School of Physics and Astronomy, Shanghai Jiao Tong University, Shanghai 200240, China}
  
  \author{Qi-An Zhang}
  \affiliation{School of Physics, Beihang University, Beijing 102206, China}

  \maketitle

\fi

\paragraph{Parameterization of the lattice-determined $\Lambda$ LCDAs.} For the convenience of phenomenological applications, we provide a simple polynomial parametrization of the lattice-determined $\Lambda$-baryon LCDAs in momentum-fraction space. The parametrization is performed in the physical momentum-fraction region where the LaMET calculation are expected to be reliable:
\begin{equation}
    x_1 \geq 0.1\ ,\qquad x_2 \geq 0.1\ ,\qquad x_3=1-x_1-x_2 \geq 0.1\ .
\end{equation}
\ifdefined\SupplementInMain
  This fit region is indicated by the interior of the green dashed triangle in Fig.~\ref{fig:final_results} of the main text.
\else
  This fit region is indicated by the interior of the green dashed triangle in Fig.~\ref{fig:final_results} of the main text~\cite{LPC:2026lcj}.
\fi
Following the convention in Appendix~E of the companion article~\cite{LPC:2026mvw}, the fitted coefficients for the first-, second-, and third-order parametrizations are given in Tables~\ref{tab:lambda_av_fit_app} and \ref{tab:lambda_t_fit_app_all} of this supplemental material, respectively. The coefficients $\widetilde{\pi}_{\Lambda,nm}$ are defined by $\widetilde{\pi}_{\Lambda,nm}=f_{\Lambda}^T/f_{\Lambda}\times \pi_{\Lambda,nm}$. 
Among these parametrizations, the second-order fit is recommended as the default input for phenomenological applications. It provides a balanced description of the LCDA shape without introducing too many weakly constrained parameters. 
The coefficient $\phi_{\Lambda,0}$ represents an overall normalization factor. 
In the present LaMET result, residual higher-power effects from the finite-momentum calculation can induce a small deviation of this factor from unity, which should be regarded as a residual systematic uncertainty in the normalization. 
For phenomenological applications, we recommend absorbing this factor into the decay constant, namely using the rescaled input $f_{\Lambda}/\phi_{\Lambda,0}$ together with the parameterized LCDAs. 
The third-order parametrization is included mainly as a benchmark and stability check, allowing one to estimate the possible impact of higher polynomial components.

The polynomial expansions used in these tables follow the convention of Ref.~\cite{Huang:2025bbk}; further details on the moment conventions are given in Appendix~E of the companion article~\cite{LPC:2026mvw}.

%%%%%%%%%%%%%%%%%%%%%%%%%%%%%%%%%%%%%%%%%%%%%%%%%%%%%%%%%%%%%%%%%%%%
\begin{table*}[htbp]
    \centering
    \caption{Fit results for $A_\Lambda$ and $V_\Lambda$ parameters at varying truncation orders, with the fitted region $x_{1,2,3} \geq 0.1$.}
    \label{tab:lambda_av_fit_app}
    \renewcommand{\arraystretch}{1.15}
    \setlength{\tabcolsep}{2pt}
    \begin{tabular}{c c c cc ccc cccc}
        \hline\hline
        Order & $\chi^2/\mathrm{dof}$ & $\phi_{\Lambda,0}$ & $\phi_{\Lambda,10}$ & $\phi_{\Lambda,11}$ & $\phi_{\Lambda,20}$ & $\phi_{\Lambda,21}$ & $\phi_{\Lambda,22}$ & $\phi_{\Lambda,30}$ & $\phi_{\Lambda,31}$ & $\phi_{\Lambda,32}$ & $\phi_{\Lambda,33}$ \\
        \hline
        $1$ & $1.026$ & $1.1242(19)$ & $0.0366(11)$ & $0.00493(36)$ & --- & --- & --- & --- & --- & --- & --- \\
        $2$ & $0.126$ & $1.1161(26)$ & $0.0343(12)$ & $0.00543(40)$ & $0.0012(14)$ & $-0.0267(18)$ & $-0.0087(14)$ & --- & --- & --- & --- \\
        $3^*$ & $0.047$ & $1.1215(29)$ & $0.0375(23)$ & $0.00565(77)$ & $-0.0005(16)$ & $-0.0215(20)$ & $-0.0055(16)$ & $0.29(11)$ & $0.007(12)$ & $0.062(13)$ & $0.171(17)$ \\
        \hline\hline
    \end{tabular}
\end{table*}
%%%%%%%%%%%%%%%%%%%%%%%%%%%%%%%%%%%%%%%%%%%%%%%%%%%%%%%%%%%%%%%%%%%%

%%%%%%%%%%%%%%%%%%%%%%%%%%%%%%%%%%%%%%%%%%%%%%%%%%%%%%%%%%%%%%%%%%%%
\begin{table}[htbp]
    \centering
    \caption{Fit results for $T_\Lambda$ parameters at varying truncation orders, with the fitted region $x_{1,2,3} \geq 0.1$.}
    \label{tab:lambda_t_fit_app_all}
    \renewcommand{\arraystretch}{1.15}
    \setlength{\tabcolsep}{6pt}
    \begin{tabular}{c c c ccc}
        \hline\hline
        Order & $\chi^2/\mathrm{dof}$ & $\widetilde{\pi}_{\Lambda,10}$ & $\widetilde{\pi}_{\Lambda,21}$ & $\widetilde{\pi}_{\Lambda,30}$ & $\widetilde{\pi}_{\Lambda,33}$ \\
        \hline
        $1$ & $0.126$ & $0.00743(10)$ & --- & --- & --- \\
        $2$ & $0.122$ & $0.00750(10)$ & $0.00043(14)$ & --- & --- \\
        $3^*$ & $0.062$ & $0.00756(21)$ & $0.00075(16)$ & $0.0180(89)$ & $-0.172(15)$ \\
        \hline\hline
    \end{tabular}
\end{table}
%%%%%%%%%%%%%%%%%%%%%%%%%%%%%%%%%%%%%%%%%%%%%%%%%%%%%%%%%%%%%%%%%%%%

% In addition to the parameterized LCDA coefficients, we also provide the normalized first moments of the quark momentum fractions in the leading-twist $\Lambda$ LCDAs. These moments summarize the average momentum sharing among the valence quarks in the lattice-determined distributions.

% \begin{table*}[htbp]
% \centering
% \caption{Normalized first moments using parameterized $\Lambda$ LCDAs in this work.}
% \label{tab:lambda_ff_f1}
% \renewcommand{\arraystretch}{1.15}
% \setlength{\tabcolsep}{6pt}
% \begin{adjustbox}{max width=\textwidth}
% \begin{tabular}{c c c c c}
% \hline\hline
% $\Lambda$ & $\langle x_1 \rangle$ & $\langle x_2 \rangle$ & $\langle x_3 \rangle$ \\
% \hline
% $A$  & $0.3273$ & $0.3273$ & $0.3455$  \\
% $V$ $(\times10^{-2})$  & $0.269$ & $-0.269$ & $0$  \\
% $T$ $(\times10^{-2})$ & $-0.178$ & $0.178$ & $0$  \\
% \hline\hline
% \end{tabular}
% \end{adjustbox}
% \end{table*}

% \begin{table*}[htbp]
% \centering
% \caption{Impacts on $g_2$ form factor in $\Lambda_b\to\Lambda\ell^+\ell^-$ at Borel mass $M_B^2=3~{\rm GeV}^2$.}
% \label{tab:lambda_ff_g2}
% \renewcommand{\arraystretch}{1.15}
% \setlength{\tabcolsep}{6pt}
% \begin{adjustbox}{max width=\textwidth}
% \begin{tabular}{c c c c}
% \hline\hline
% $g_2$ & Asymptotic form & This work with $\phi_{\Lambda,0}=1$ & This work with $\phi_{\Lambda,0}=1.1161$ \\
% \hline
%  & $-0.0236374$ & $-0.0231934$ & $-0.0224884$ \\
% \hline\hline
% \end{tabular}
% \end{adjustbox}
% \end{table*}

\ifdefined\SupplementInMain
  % Bibliography is handled by the main arXiv file.
\else
  \bibliography{ref}
  \end{document}
\fi
%%%%%%%%%%%%%%%%%%%%%%%%%%%%%%%%%%%%%%%%%%%%%%%%%%%%%%%%%%%%%%%%%%%%

\end{document}